\documentclass[a4paper, 11pt]{article}
\usepackage{lineno,hyperref}
\modulolinenumbers[1]
\usepackage[OT1]{fontenc} 
\usepackage{times}

\usepackage{graphicx}

\usepackage{algorithmic}%
\usepackage{amsmath,amssymb}
\usepackage{amsthm}
\usepackage{multirow}
\usepackage{array}
\usepackage{color}
\usepackage[table]{xcolor}
\usepackage[bottom]{footmisc}
\usepackage{subfig}
\usepackage{url}

\usepackage{rotating}
\usepackage{multirow}
\usepackage{subfig}
\usepackage[final]{pdfpages}
\usepackage{chngpage}
\usepackage{rotating}
\newcommand{\shrink}[1]{}

\usepackage[noresetcount,algoruled,noend,norelsize,vlined]{algorithm2e}

\def\X{\mathcal{X}}

\SetKwFunction{A}{\cal A}

\SetKwFunction{updgvns}{UPDGVNS}
\SetKwFunction{udgvns}{UDGVNS}
\SetKwFunction{dgvns}{DGVNS}
\SetKwFunction{dfbb}{DFBB}
\SetKwFunction{ilds}{ILDS}
\SetKwFunction{btd}{BTD-HBFS}
\SetKwFunction{vnsalgo}{VNS/LDS+CP}

\SetKwFunction{lds}{LDS}
\SetKwFunction{ldsr}{LDS$^r$}
\SetKwFunction{lb}{lb}
\SetKwFunction{pop}{pop}
\SetKwFunction{push}{push}

\SetKwFunction{master}{master}
\SetKwFunction{worker}{worker}
\SetKwFunction{fneighbor}{NH}
\SetKwFunction{incl}{$+_\ell$}
\SetKwFunction{inck}{$+_k$}
\SetKwFunction{inc}{$+$}
\SetKwFunction{mult}{mult2}
\SetKwFunction{add}{add1}
\SetKwFunction{addjump}{add1/jump}
\SetKwFunction{luby}{Luby}
\SetKwFor{Procedure}{Procedure}{}{}%
\SetKwFor{Function}{Function}{}{}%
\SetKwFunction{send}{Send}
\SetKwFunction{receive}{Receive}

\SetKwFunction{Hvoisinage}{Random}
\SetKwFunction{Complete}{CompleteCluster}
\SetKwFunction{best}{BestSol}
\SetKwFunction{update}{Update}
\SetKwFunction{getNgh}{getNeighborhood}
\SetKwFunction{Conflict}{getConflict}

\begin{document}
		
		\title{Computation Protein Design instances with small tree-width:  selection based on coarse approximated  3D average position volume  }

\author{David Allouche \\ 
INRA, MIA Toulouse, UR-875, 31320 Castanet-Tolosan, France 
\\david.allouche@inra.fr}


\maketitle
 \section*{Keywords:}
			benchmark,
			Computational protein Design,
			rosetta force-field,
			coarse structural metric,
			dumbrack rotamer library,
			Graphical model, 
			Most probable explanation (MPE) problem, 
			Cost function network, 
			Tree decomposition. 
		
\section*{Abstract}
This paper proposes small tree-width graph decomposition  computational protein design CFN instances defined according to the model ~\cite{allouche2012computational} with protocol  defined  by Simononcini et al ~\cite{Simoncini2015GuaranteedProblems} .
The proteins used in the benchmark have been selected in the PDB (not on their biological interest) to explore the efficiency of global search method based on tree-width decomposition.  The instances are bigger than those  previously proposed in  the paper ~\cite{Simoncini2015GuaranteedProblems} with one backbone relaxation and the aka Beta November 2016  Rosetta force-field \cite{Alford106054}. The benchmark includes 21 proteins selected with a low level of sequences identity (40\%) . Those instances have been selected on the basis of 3D criteria by applying a decreasing average coarse  volume occupancy filter by Amino Acid (-i.e. by CFN variable) .
The instances characteristic (see Table~\ref{pdb:instances}) contain from $130$ up to $n = 282$ variables with a maximum domain size from $383$ to $438$, and between $1706$ and $6208$ cost functions. The min-fill tree-width ranges from $21$ to $68$, and from $0.16$ to $0.34$ for a normalized tree width.
Those instances have been used  for   UDGVNS search algorithm\cite{Ouali17} benchmarking. This approach is suitable for evaluation of search methods that exploit the notion of graph decomposition.

\begin{table}[t]
	\centering
	\scalebox{0.9}{
		\begin{tabular}{c|c|c|c|c|c||c|c|c}
			\hline
			pdbid   &       $|X|$   &       $d$     &       $e$     &       $tw$            & $tw/|X|$      &$\min(Rg(i)/Rg)$ &  $\bar{V} (\AA^3/var)$ & $|S|= \log_{10} \prod_{D_i}$     \\
			\hline
			5dbl    &       130     &       384     &       1,706    &       21              & 0.16          &       0.150   &     1,212.49		& 303.3 \\ 
			5jdd    &       263     &       406     &       5,220    &       41              & 0.16          &       0.239   &       655.58		& 627.4 \\ 
			3r8q    &       271     &       418     &       5,518    &       43              & 0.16          &       0.341   &       472.88		& 640.3 \\ 
			4bxp    &       170     &       439     &       2,636    &       33              & 0.19          &       0.316   &       457.81		& 402.1 \\ 
			1f00    &       282     &       430     &       6,208    &       51              & 0.18          &       0.269   &       439.28 	& 660.1 \\ 
			2x8x    &       235     &       407     &       4,745    &       44              & 0.19          &       0.354   &       404.42 	& 559.0 \\ 
			1xaw    &       107     &       412     &       1,623    &       28              & 0.26          &       0.308   &       378.04 	& 259.5 \\ 
			5e10    &       133     &       400     &       2,286    &       34              & 0.26          &       0.294   &       344.68 	& 310.8 \\ 
			1dvo    &       152     &       389     &       2,587    &       51              & 0.34          &       0.420   &       343.38 	& 361.6 \\			
			1ytq    &       181     &       415     &       3,449    &       54              & 0.30          &       0.392   &       332.69 	& 422.2 \\ 
			2af5    &       292     &       410     &       5,693    &       68              & 0.23          &       0.427   &       330.23 	& 686.2 \\			
			1ng2    &       176     &       397     &       3,135    &       60              & 0.34          &       0.473   &       309.16 	& 414.9 \\ 
			3sz7    &   	151 	&       450     &       2805     &       49              & 0.32          &       0.403   &		 304.87 	& 355.5 \\
			2gee    &       188     &       397     &       3,715    &       38              & 0.20          &       0.367   &       293.25 	& 445.2 \\ 
			5e0z    &       136     &       420     &       2,367    &       36              & 0.26          &       0.362   &       279.00 	& 316.0 \\ 
			1yz7    &       176     &       418     &       3,538    &       49              & 0.28          &       0.414   &       276.35 	& 419.3 \\ 			
			3e3v    &       154     &       436     &       2,976    &       37              & 0.26          &       0.367   &       251.97 	& 368.3 \\ 
			3lf9    &       120     &       416     &       2,133    &       31              & 0.24          &       0.323   &       251.51		& 286.3	\\ 
			1is1    &       185     &       431     &       3,740    &       48              & 0.26          & 		 0.459   &	 	 245.58  	& 443.1 \\
			5eqz    &       138     &       434     &       2,567    &       33              & 0.24          &       0.338   &       241.93 	& 330.5 \\ 
			4uos    &       188     &       383     &       4,161    &       44              & 0.23          &       0.347   &       234.11 	& 455.8 \\ 
			\hline
		\end{tabular}
		}
	\vspace*{-.2cm}
	\caption{Characteristics of PDB instances:
		pdbid is the code reference in PDB database, $|\X|$ is the number of variables, $d$ is the maximum domain size, $e$ is the number of cost functions, $tw$ is the min-fill tree-width and  $tw/|\X|$ a normalized tree-width by $|\X|$. The last tree columns correspond to structural criteria respectively defined in  (eq: \ref{rgVol}) , (eq: \ref{minRgRate}) and the log of the domain Cartesian product.  	$\bar{V}$ is used for PDB list order.
	}
	\label{pdb:instances}
\end{table}

\newpage

\section{CPD background}
Structure-based computational protein design (CPD) plays a critical role in advancing the field of protein engineering.
In the past decade the field  has rapidly expanded, providing an approach to test the structural basis for function as well as a tool for designing useful molecules ~\cite{ReviewLIPPOW2007305}~\cite{Reviews:huang2016coming}~\cite{bakerDenovoReview2019}.

 Using an all-atom energy function, CPD tries to identify amino acid sequences that fold into a target structure and ultimately perform a desired function.
The CPD problem is the inverse problem of the protein folding~\cite{Pabo1983} ( The fold is known),  it aims to find the best sequences and side chain conformation as to minimize the total energy of the system . 

The Total Energy is reformulated as follows~\cite{Dahiyat1996}:
\begin{equation}\label{eq-energy}
E_T = E_\varnothing + \sum_i E(i_r) +\sum_i \sum_{j>i} E(i_r,j_s)
\end{equation}

where $E_\varnothing$ , $E(i_r)$ , $E(i_r,j_s)$ are respectively the backbone template , internal side chain , and side-chain pairwise energy terms calculated from protein force fields (such as CHARMM , AMBER or rosetta).

The CPD is then formulated with the goal of identifying a
conformation of minimum energy via the mutation of a specific
subset of amino acid residues, \emph{i.e.} by affecting their identity and
their 3D orientations (rotamers).  The conformation that minimizes the energy is called the \emph{GMEC} (Global Minimum Energy Conformation). The GMEC corresponds to a maximum probability mass due to the Boltzmann relation between molecular energy and probability, which is equivalent to the Markov Random Field modeling with a Maximum A Posteriori probability (MAP-MRF) estimation. The GMEC search is NP-hard\cite{Pierce2002} and has be formulated  as a  Cost  Function Network  (CFN)~\cite{allouche2012computational}. Compared to other complete formulations, the CNF modeling is the state of the art model for GMEC resolution ~\cite{CPD-AIJ}.
%

\section{The Computational Protein Design challenge}

Computational Protein Design faces several challenges. The exponential size of the conformational and protein
sequence space that has to be explored rapidly grows out of the reach of computational approaches. Another obstacle to overcome is the accurate structure prediction for a given sequence~\cite{khoury2014,gront2011}. Therefore and in order to reduce the problem to the identification of an amino acid sequence that can fold into a a target 3D-scaffold that matches the design objective ~\cite{bowie1991}, the design problem is usually approached as an inverse folding problem~\cite{Pabo1983}.  In structural biology, the stability of a conformation can be directly evaluated through the energy of the conformation, a stable fold being of minimum energy~\cite{Anfinsen73}.

In CPD, two approximations are common. First, it is assumed that the resulting designed protein retains the overall fold of the chosen
scaffold: the protein \emph{backbone} is considered fixed. At specific positions chosen automatically or by the molecular modeler, the amino acid used can be modified, thus changing the \emph{side chain} . 
Second, the domain of conformations available to each amino acid side
chain is continuous. 

\begin{figure}
	\centerline{\includegraphics[width=0.4 \textwidth]{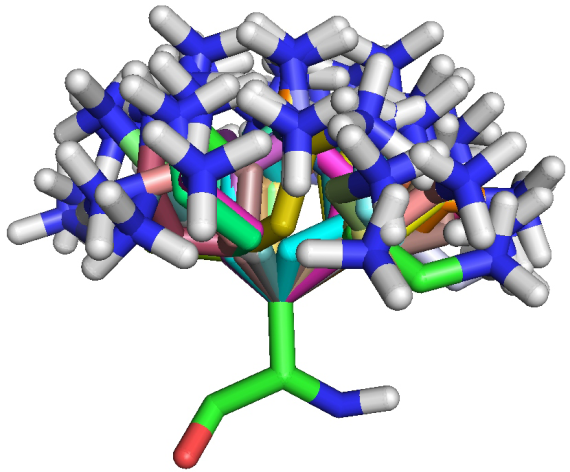}}
	\caption{
		example of side chain rotamer set of the lysine Amino Acid
		\label{rotpng}}
\end{figure} 

This continuous domain is approximated using a set of discrete conformations defined by the value of their inner dihedral angles. These conformations, or \emph{rotamers}~\cite{janin1978}, are derived from the most frequent conformations in the experimental repository of known protein structures, the PDB (Protein Data Bank, \url{www.pdb.org}). Different rotamer libraries dumbrack ~\cite{Campeotto13} , penultimate ~\cite{Lovell2000} and tuffery ~\cite{Tuffery1991AConformations}  have been used in constraint-based approaches for GMEC search for protein design.\cite{allouche2012computational},\cite{Simoncini2015GuaranteedProblems}, \cite{Allouchejcmi2018a}.

\begin{figure}
	\centerline{\includegraphics[width=1.0\textwidth]{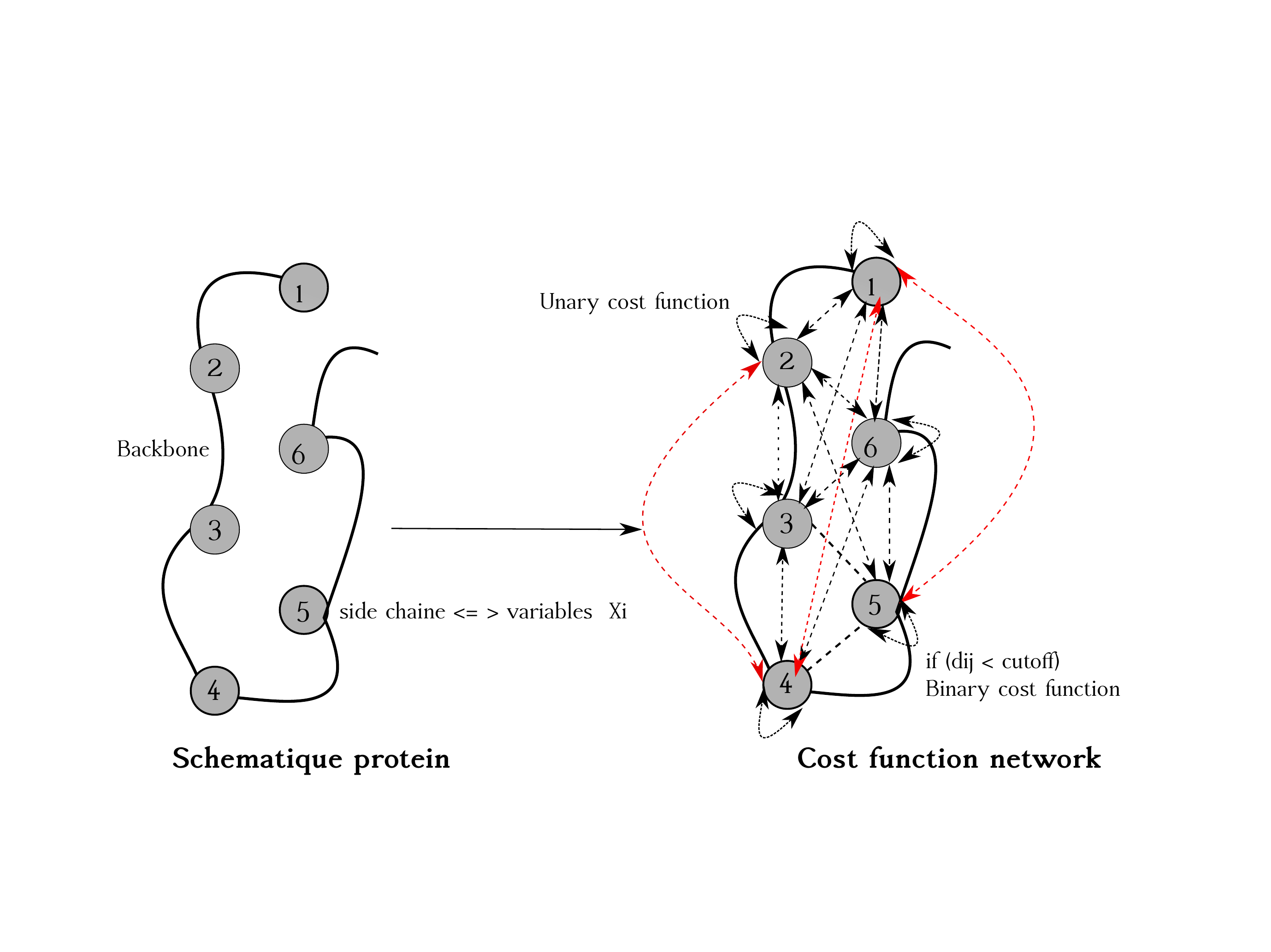}}
	\caption{
		schematic view of cfn modeling for single backbone protein design:
		each position correspond to cfn variable. when position distance are lower than the cutoff distance use by the force-field a binary constraint is  added in the instances . note that due to this cutoff distance a spherical shape and  cutoff distance used in the Energy calculation, non Gobular protein will induce constraint graph sparcity and decomposable cfn instances and spherical protein will be clause from click. 
		\label{fig1}}
\end{figure} 

\section{CFN model definition}
CPD instances can be directly represented as Cost Function Networks (CFN). 
	which is a pair $(X,W)$ where $X = \{1,
	\ldots, n\}$ is a set of $n$ variables and $W$ is a set of cost
	functions.  Each variable $i \in X$ has a finite domain $D_i$ of
	values that can be assigned to it. A value $r\in D_i$ is denoted
	$i_r$. For a set of variables $S\subseteq X$, $D_S$ denotes the
	Cartesian product of the domains of the variables in $S$. For a given
	tuple of values $t$, $t[S]$ denotes the projection of $t$ over $S$. A
	cost function $w_S \in W$, with scope $S\subseteq X$, is a function
	$w_S: D_S \mapsto [0,k]$ where $k$ is a maximum integer cost used for
	forbidden assignments.

We assume, without loss of generality, that every CFN includes at
least one unary cost function $w_i$ per variable $i\in X$ and a
nullary cost function $w_\varnothing$. All costs being non-negative,
the value of this constant function, $w_\varnothing$, provides a lower bound on the cost of any assignment.

The Weighted Constraint Satisfaction Problem (WCSP) is to find a
complete assignment $t$ minimizing the combined cost function
$\bigoplus_{w_S \in W} w_S(t[S])$, where $a \oplus b = \min(k, a + b)$ is the $k$-bounded addition. This optimization problem has an
associated NP-complete decision problem. Notice that if $k=1$, then the WCSP is nothing but the famous Constraint Satisfaction Problem or CSP (not the Max-CSP).

Modeling the CPD problem as a CFN is straightforward. The set of
variables $X$ has one variable $i$ per residue $i$. The domain of each
variable is the set of \textit{(amino acid,conformation)} pairs in the
rotamer library used. The global energy function can be represented by
0-ary, unary and binary cost functions, capturing the
constant energy term $w_\varnothing=E_\varnothing$, the unary energy
terms $w_i(r)=E(i_r)$, and the binary energy terms $w_{ij}(r,s) =
E(i_r, j_s)$, respectively. 


Notice that there is one discrepancy between the original formulation
and the CFN model: Energies are represented as arbitrary floating
point numbers while CFN uses positive costs. This can simply be fixed
by first subtracting the minimum energy from all energy factors. These
positive costs can then be multiplied by a large integer constant $M$
and rounded to the nearest integer if integer costs are required.

The first the CPD problem was introduced with rigid backbone. The problem can be naturally
expressed as a Cost Function Network (CFN) and solved as a Weighted Constraint Satisfaction Problem ~\cite{allouche2012computational}.

\section{The data source:}

Proteins are one of the most versatile modular assembling systems in nature. Experimentally,
more than 127 000 protein structures have been identified and more are deposited every day
in the Protein Data Bank. The Protein Data Bank (PDB)~\cite{pdb}) is a database for the three-dimensional structural data of large biological molecules, such as proteins and nucleic acids. The data, typically obtained by X-ray crystallography, NMR spectroscopy, or, increasingly, cryo-electron microscopy, and submitted by biologists and biochemists from around the world, are freely accessible on the Internet via the websites.

\begin{figure}
	\centering
	\includegraphics[width=1.0\textwidth]{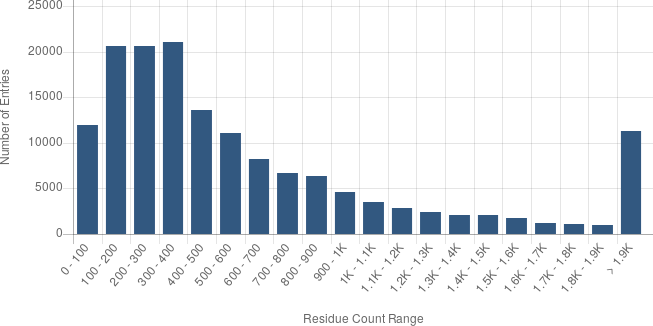}
	\caption{A representation of distribution of the number of residues by entries in the pdb. More statistics about  entries  can be follow 	at \url{https://www.rcsb.org/stats/}.}
	\label{fig:pdbsize}
\end{figure}
\begin{figure}
	\centering
	\includegraphics[width=0.7\textwidth]{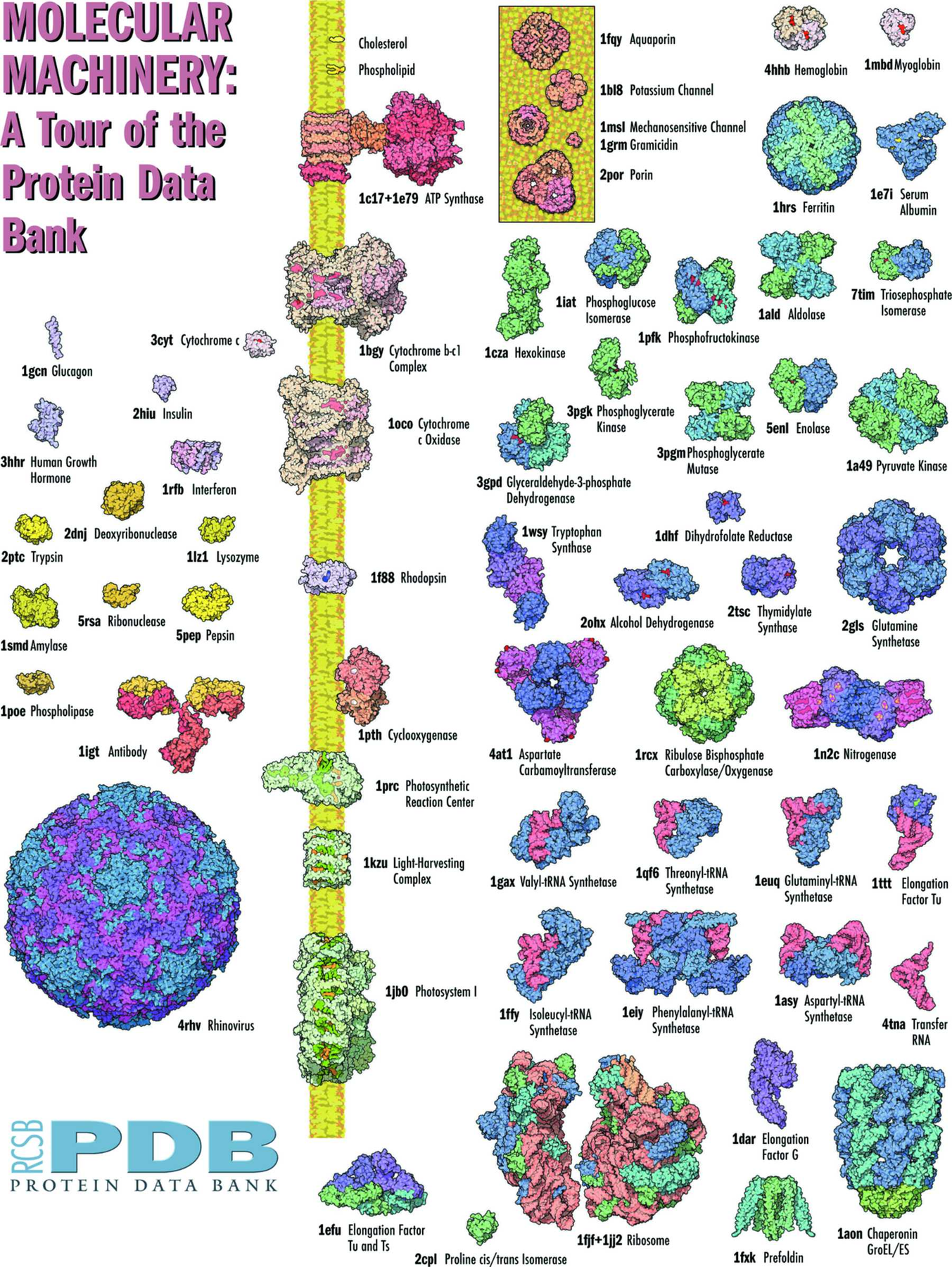}
	\caption{A look at the diversity of structures in the PDB archive. These images, shown to scale, were created by David S. Goodsell (The Scripps Research Institute).  Original figure from Berman et Al \cite{PDBinfo:Berman:sc5004})
		An expanded version of this figure is available for download from http://www.pdb.org}
	\label{fig:pdbfold}
\end{figure}

In the past, the number of structures in the PDB has grown at an approximately exponential rate, passing the 100 registered structures milestone in 1982, 1,000 in 1993, 10,000 in 1999 and 100,000 in 2014. However, since 2007  the rate of accumulation of new protein structures appears to have plateaued ~\cite{PDBinfo:Berman:sc5004}. Entries sizes distribution can be shown  the histogram \ref{fig:pdbsize}. The database contains all the known fold, an overview of which can be seen on figure \ref{fig:pdbfold}. 

%
%

\section{Instances selection protocol}

In this report we would like to evaluate CFN search  capability to solve difficult structured instances  in terms of constraint graph decomposition. Accordingly, we tried to generate new larger instances than those previously generated in  \cite{Simoncini2015GuaranteedProblems}. Furthermore we used as ordering criteria $\bar{V}$\ref{3D_CRIT}  in order to exhibit instances with small tree-width . This coarse metric is an heuristic for filtering  constraint graph sparsity  based on a 3D criteria.   

Due to the huge number of entries,  a pre-selection set has to be done in order to extract subset of protein of interest for benchmarking. The NP-hard resolution of CPD instances requires the selection of a small number of problems for further experimentation. 
In Simoncini and al\cite{Simoncini2015GuaranteedProblems} the benchmark set was extracted from the PDB and filtered with the following criteria: monomeric proteins with an X-ray resolved structure below 2~\AA, with no missing or non standard residues and no ligand. Chain length was limited to 100 amino-acids. A total of 107 proteins were extracted as of the $1^{st}$ of September 2014, retrieving only representative structures at 30\% sequence identity. The chain lengths scale from 50 to 100 residues, defining a
collection of problems of gradually increasing complexity. Each protein was then relaxed 10 times with the default Rosetta relax protocol~\cite{LeaverFay2011}, using Talaris2014 energy
function~\cite{talaris2014} and the backbone of lowest energy used for benchmarking (See  SI  for a detailed list of the proteins~\cite{Simoncini2015GuaranteedProblems} ).

\subsection{PDB query and model}
 In this work,  each protein structure was fully redesigned  in a similar way to the~\cite{Simoncini2015GuaranteedProblems} protocol. On the basis of energy matrix generated with a modified release of {\sc pyrosetta.4} script \cite{Simoncini2015GuaranteedProblems} but with a single backbone relaxation and the BetaNov16 Rosetta forcefields .
 
The current benchmark has been selected from PDB (release Jan. 2017). The protein sizes range between $100$ and $300$ amino-acids. 

The PDB query has been filtered with the following additional criteria: Resolution has to be lower than $2.5$ \AA; membrane proteins, protein complexes, as well as proteins with disulfur bridges have been removed; in addition, proteins including non natural amino acid have also been removed. We also discarded proteins with missing residues, out of the N and C terminal part of the sequence in order to select a protein subset without any hole. Proteins with identity sequences higher than $40\%$ were not selected. 

The corresponding remaining set includes 436 PDB references. The full list of the corresponding pdb identifiers is available in the supplementary spreadsheet document . It was furthermore sorted by the 3D coarse criteria describes below.  

\subsection{3D critter filtering}\label{3D_CRIT}

\begin{figure}
	\centering
	\includegraphics[width=0.7\linewidth]{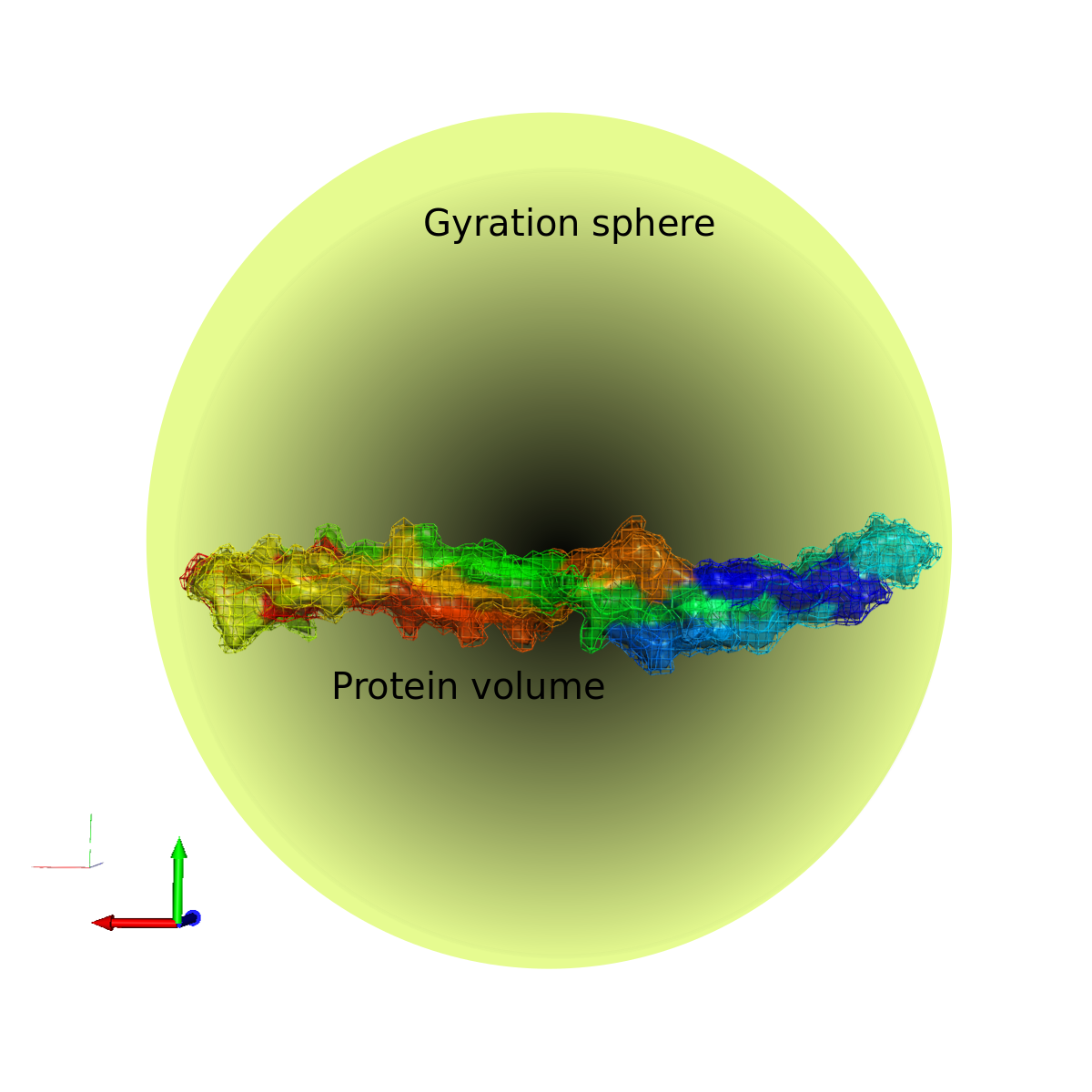}
	\caption{ryration radius representation on a given pdb structure}
	\label{fig:sphere_rad}
\end{figure}

In order to select well decomposable instances, after preliminary reorientation of each protein according to its inertial moment we filtered the resulting PDB set by an heuristic criteria based on a coarse estimation of the protein volume based on the gyration radius (eq: \ref{Rg}). The coarse volume obtained from the gyration radius is correlated to the number of amino-acids and accordingly to the number of variables in the model.

After  normalisation,  the estimated volume $\bar{V}$ represents a coarse average volume by position. \ref{rgVol}.

 An example of gyration sphere is represented in figure : \ref{fig:sphere_rad}.
 
\begin{equation}\label{Rg}
Rg =  \sqrt{\sum_{i}^{n}\frac{{(\bar{x}-x_i)^2+(\bar{y}-y_i)^2+(\bar{z}-z_i)^2}}{n}}
\end{equation}
\begin{equation} \label{rgVol}
\bar{V}=\frac{\frac{4}{3}.\pi.Rg^3}{|X|}
\end{equation}

The  decreasing sort based on $\bar{V}$ gave us a protein ordering from the least to the most spherical fold.

 By  construction and due to the cut-off distance used for  energy calculations and its related CFN,  globular proteins \footnote{Globular proteins or spheroproteins are spherical ("globe-like") proteins \url{https://en.wikipedia.org/wiki/Globular_protein}} will closely correspond to clique.  Non spherical proteins however induce constraint graph sparsity and a well decomposed graph of constraints. 
 
Smaller $\bar{V}$ values correspond to  proteins closer to a spherical fold that includes less free space inside the sphere based on the gyration radius. 

\subsection{Alternative 3D criteria:}

It should be noted that other  alternative filters can be used for instance ordering. Thus, the spherical shape can be detected by similar components  values Rg(i) of the radius of gyration (eq: \ref{Rg}). Where Rg($x$) (eq:\ref{Rgx}) , Rg($y$) (eq:\ref{Rgy}) and Rg($z$) (eq:\ref{Rgz}) correspond respectively to sub components defined as the root mean square distance from  all atoms position to their centroid around the axis $x$ , $y$ , $z$ 
-i.e namely  gyration radius component in the  plan  $yz$ , $xz$ ,$xy$  (orthogonal to the  axis ($x$,$y$,$z$)).

\begin{equation}\label{Rgx}
Rg(x)=  \sqrt{\sum_{i}^{n}\frac{{(\bar{y}-y_i)^2+(\bar{z}-z_i)^2}}{n}}
\end{equation}
\begin{equation}\label{Rgy}
Rg(y)=  \sqrt{\sum_{i}^{n}\frac{{(\bar{x}-x_i)^2+(\bar{z}-z_i)^2}}{n}}
\end{equation}
\begin{equation}\label{Rgz}
Rg(z)=  \sqrt{\sum_{i}^{n}\frac{{(\bar{x}-x_i)^2+(\bar{y}-y_i)^2}}{n}}
\end{equation}

\begin{equation} \label{minRgRate}
min(Rg(x)/Rg, Rg(y)/Rg, Rg(z)/Rg) 
\end{equation}

For selecting non spherical protein, an alternative way consists in  first calculating the $Rg(i)$, and then filtering the proteins  by the new criteria (eq:\ref{minRgRate}).

because , the  gyration component refers to the distribution of the atoms of an 3D structure around associated axis.

For  spheroid proteins , all components  Rg(x) , Rg(y), Rg(z) and Rg are asymptotically equal . therefore  $(Rg(i)/Rg)$ is close to 1. For non spherical fold this ratio is far from 1 as is the $min_i(Rg(i)/Rg)$ (eq:\ref{minRgRate}). 

Consequently this  criteria is an other method to coarsely detect the non spheroid shape , when  gyration radius are dissimilar in the plan (X,Y) (Y,Z) and (Z,X) due to difference observed  in the components Rg(x) , Rg(y),  Rg(z).

An increasing sort of the PDB list based on those criteria (eq: ~\ref{minRgRate} ) produces a new protein ordering form the least to the most spherical. Compared to the previous $\bar{V}$ ordering , this new criteria produces the same 18 first instances set with  re-ranking (except the 2 first proteins).  



\section{ Conclusion}

In this work we present  two coarse  structural criterions  for  small tree-width CPD instances filtering, both of which are heuristic. Our goal is not to calculate the exact volume but to detect  non globular -i.e non spheroids- characters due to their putative highly decomposable properties .

Arbitrarily we used the  instances  done with the $\bar{V}$ \ref{rgVol} associated to coarse  average volume per variable. From the 436 putative instances resulting form the PDB query, we extracted only the 21 first elements and used them as our benchmark set. The instances characteristics (see Table~\ref{pdb:instances}) contain from $130$ up to $n = 282$ variables with a maximum domain size from $383$ to $438$, and between $1706$ and $6208$ cost functions. The min-fill tree-width ranges from $21$ to $68$, and from $0.16$ to $0.34$ for a normalized tree-width.
(See  SI  for all instances detailed list of the proteins in the attached spreadsheet ). 

Note that those instances have been ; with others; used  in updgvns algorithm  benchmark. The method is a  Variable Neighborhood Search method for CFN resolution \cite{ouali2017}.

\section*{References}

\bibliography{CPD,ijcai17}

\begin{thebibliography}{10}
\expandafter\ifx\csname url\endcsname\relax
  \def\url#1{\texttt{#1}}\fi
\expandafter\ifx\csname urlprefix\endcsname\relax\def\urlprefix{URL }\fi
\expandafter\ifx\csname href\endcsname\relax
  \def\href#1#2{#2} \def\path#1{#1}\fi

\bibitem{allouche2012computational}
D.~Allouche, et~al., {Computational protein design as a cost function network
  optimization problem}, in: Proc.\ of CP, 2012, pp. 840--849.

\bibitem{Simoncini2015GuaranteedProblems}
D.~Simoncini, D.~Allouche, S.~de~Givry, C.~Delmas, S.~Barbe, T.~Schiex,
  Guaranteed discrete energy optimization on large protein design problems, J.
  of Chemical Theo. and Comput. 11(12) (2015) 5980--5989.

\bibitem{Alford106054}
R.~Alford, et~al., The rosetta all-atom energy function for macromolecular
  modeling and design, Journal of Chemical Theory and Computation 13~(6) (2017)
  3031--3048.

\bibitem{Ouali17}
A.~Ouali, D.~Allouche, S.~de~Givry, S.~Loudni, Y.~Lebbah, F.~Eckhardt,
  L.~Loukil, {Iterative Decomposition Guided Variable Neighborhood Search for
  Graphical Model Energy Minimization}, in: Proc.\ of UAI-17, Sydney,
  Australia, 2017, pp. 550--559.

\bibitem{ReviewLIPPOW2007305}
S.~M. Lippow, B.~Tidor,
  \href{http://www.sciencedirect.com/science/article/pii/S0958166907000778}{Progress
  in computational protein design}, Current Opinion in Biotechnology 18~(4)
  (2007) 305 -- 311, protein technologies / Systems biology.
\newblock \href
  {http://dx.doi.org/https://doi.org/10.1016/j.copbio.2007.04.009}
  {\path{doi:https://doi.org/10.1016/j.copbio.2007.04.009}}.
\newline\urlprefix\url{http://www.sciencedirect.com/science/article/pii/S0958166907000778}

\bibitem{Reviews:huang2016coming}
P.-S. Huang, S.~E. Boyken, D.~Baker, The coming of age of de novo protein
  design, Nature 537 (2016) 320--327.

\bibitem{bakerDenovoReview2019}
D.~Baker, \href{https://onlinelibrary.wiley.com/doi/abs/10.1002/pro.3588}{What
  has de novo protein design taught us about protein folding and biophysics?},
  Protein Science 28~(4) (2019) 678--683.
\newblock \href
  {http://arxiv.org/abs/https://onlinelibrary.wiley.com/doi/pdf/10.1002/pro.3588}
  {\path{arXiv:https://onlinelibrary.wiley.com/doi/pdf/10.1002/pro.3588}},
  \href {http://dx.doi.org/10.1002/pro.3588} {\path{doi:10.1002/pro.3588}}.
\newline\urlprefix\url{https://onlinelibrary.wiley.com/doi/abs/10.1002/pro.3588}

\bibitem{Pabo1983}
C.~Pabo, {Molecular technology. Designing proteins and peptides.}, Nature
  301~(5897) (1983) 200.

\bibitem{Dahiyat1996}
B.~I. Dahiyat, S.~L. Mayo, {Protein design automation.}, Protein science 5~(5)
  (1996) 895--903.

\bibitem{Pierce2002}
N.~A. Pierce, E.~Winfree, {Protein design is NP-hard.}, Protein engineering
  15~(10) (2002) 779--82.

\bibitem{CPD-AIJ}
D.~Allouche, I.~Andr{\'e}, S.~Barbe, J.~Davies, S.~de~Givry, G.~Katsirelos,
  B.~O'Sullivan, S.~Prestwich, T.~Schiex, S.~Traor{\'e}, Computational protein
  design as an optimization problem, Artificial Intelligence 212 (2014) 59--79.

\bibitem{khoury2014}
G.~A. Khoury, J.~Smadbeck, C.~A. Kieslich, C.~A. Floudas, Protein folding and
  \emph{de novo} protein design for biotechnological applications, Trends in
  biotechnology 32~(2) (2014) 99--109.

\bibitem{gront2011}
D.~Gront, D.~W. Kulp, R.~M. Vernon, C.~E. Strauss, D.~Baker, Generalized
  fragment picking in rosetta: design, protocols and applications, PloS one
  6~(8) (2011) e23294.

\bibitem{bowie1991}
J.~U. Bowie, R.~Luthy, D.~Eisenberg, A method to identify protein sequences
  that fold into a known three-dimensional structure, Science 253~(5016) (1991)
  164--170.

\bibitem{Anfinsen73}
C.~Anfinsen, Principles that govern the folding of protein chains, Science
  181~(4096) (1973) 223--253.

\bibitem{janin1978}
J.~Janin, S.~Wodak, M.~Levitt, B.~Maigret, Conformation of amino acid
  side-chains in proteins, Journal of molecular biology 125~(3) (1978)
  357--386.

\bibitem{Campeotto13}
F.~Campeotto, A.~Dal~Palù, A.~Dovier, F.~Fioretto, E.~Pontelli, A constraint
  solver for flexible protein models, J. Artif. Int. Res. (JAIR) 48~(1) (2013)
  953--1000.

\bibitem{Lovell2000}
S.~C. Lovell, J.~M. Word, J.~S. Richardson, D.~C. Richardson, {The penultimate
  rotamer library.}, Proteins 40~(3) (2000) 389--408.

\bibitem{Tuffery1991AConformations}
P.~Tuffery, C.~Etchebest, S.~Hazout, R.~Lavery, {A new approach to the rapid
  determination of protein side chain conformations.}, Journal of biomolecular
  structure {\&} dynamics 8~(6) (1991) 1267--89.

\bibitem{Allouchejcmi2018a}
A.~Charpentier, D.~Mignon, S.~Barbe, J.~Cortes, T.~Schiex, T.~Simonson,
  D.~Allouche, Variable neighborhood search with cost function networks to
  solve large computational protein design problems, Journal of Chemical
  Information and Modeling 59~(1) (2019) 127--136.

\bibitem{pdb}
H.~M. Berman, J.~Westbrook, Z.~Feng, G.~Gilliland, T.~Bhat, H.~Weissig, I.~N.
  Shindyalov, P.~E. Bourne, The protein data bank, Nucleic acids research
  28~(1) (2000) 235--242.

\bibitem{PDBinfo:Berman:sc5004}
H.~M. Berman, \href{https://doi.org/10.1107/S0108767307035623}{{The Protein
  Data Bank: a historical perspective}}, Acta Crystallographica Section A
  64~(1) (2008) 88--95.
\newblock \href {http://dx.doi.org/10.1107/S0108767307035623}
  {\path{doi:10.1107/S0108767307035623}}.
\newline\urlprefix\url{https://doi.org/10.1107/S0108767307035623}

\bibitem{LeaverFay2011}
A.~Leaver-Fay, M.~Tyka, S.~M. Lewis, O.~F. Lange, J.~Thompson, R.~Jacak,
  K.~Kaufman, P.~D. Renfrew, C.~A. Smith, W.~Sheffler, I.~W. Davis, S.~Cooper,
  A.~Treuille, D.~J. Mandell, F.~Richter, Y.-E.~A. Ban, S.~J. Fleishman, J.~E.
  Corn, D.~E. Kim, S.~Lyskov, M.~Berrondo, S.~Mentzer, Z.~Popovi{\'c}, J.~J.
  Havranek, J.~Karanicolas, R.~Das, J.~Meiler, T.~Kortemme, J.~J. Gray,
  B.~Kuhlman, D.~Baker, P.~Bradley, Rosetta3: an object-oriented software suite
  for the simulation and design of macromolecules., Methods Enzymol. 487 (2011)
  545--574.

\bibitem{talaris2014}
M.~J. O'Meara, A.~Leaver-Fay, M.~Tyka, A.~Stein, K.~Houlihan, F.~DiMaio,
  P.~Bradley, T.~Kortemme, D.~Baker, J.~Snoeyink, B.~Kuhlman, A combined
  covalent-electrostatic model of hydrogen bonding improves structure
  prediction with rosetta, J. Chem. Theory Comput. 11~(2) (2015) 609--622.

\bibitem{ouali2017}
A.~Ouali, D.~Allouche, S.~De~Givry, S.~Loudni, Y.~Lebbah, F.~Eckhardt,
  L.~Loukil, Iterative decomposition guided variable neighborhood search for
  graphical model energy minimization, in: Conference on Uncertainty in
  Artificial Intelligence, UAI 2017, 2017.

\end{thebibliography}
\end{document}